\documentclass[12pt]{article}

\usepackage{scicite}

\usepackage{times}
\usepackage{graphicx}
\usepackage{chemformula}
\usepackage{multirow} 
\usepackage{amsmath}
\usepackage{float}
\usepackage{arydshln}

\newenvironment{sciabstract}{%
\begin{quote} \bf}
{\end{quote}}

\title{Substrate stiffness modulates bacterial adhesion and diversity of adherent phenotypes across growth stages}

\author
{René Riedel$^{1}$, Garima Rani$^{1}$, Anupam Sengupta $^{1,2^\ast}$\\
\\
\normalsize{$^{1}$Physics of Living Matter Group, Department of Physics and Materials Science,}\\
\normalsize{University of Luxembourg, 162 A, Avenue de la Fa\"{i}encerie, L-1511, Luxembourg}\\
\normalsize{$^{2}$Institute for Advanced Studies, University of Luxembourg,}\\
\normalsize{2, Avenue de l’Université, L-4365, Esch-sur-Alzette, Luxembourg}\\
\normalsize{$^\ast$E-mail: anupam.sengupta@uni.lu}
}

\date{}

\begin{document} 


\baselineskip24pt


\maketitle


\begin{sciabstract}

Surface-adhesion and stiffness of underlying substrates mediate geometry, mechanics and self-organization of expanding bacterial colonies. Recent studies have qualitatively indicted that stiffness may impact bacterial attachment and accumulation, yet the variation of cell-to-surface adhesion with substrate stiffness remains to be quantified. Here, by developing a cell-level Force Distance Spectroscopy (FDS) technique based on Atomic Force Microscopy (AFM), we simultaneously quantify the cell-surface adhesion alongside stiffness of the underlying substrates to reveal stiffness-dependent adhesion in phototrophic bacterium \textit{Chromatium okenii}. As stiffness of the soft substrate, modelled via low-melting-point (LMP) agarose pad, was varied between 20 kPa and 120 kPa by changing agarose concentrations, we observe a progressive increase of the mean adhesion force by over an order of magnitude, from $0.21\pm0.10$ nN to $2.42\pm1.16$ nN. In contrast, passive polystyrene (PS) microparticles of comparable dimensions showed no perceptible change in their surface adhesion, confirming that the stiffness-dependent adhesive interaction of \textit{C. okenii} is of biological origin. Furthermore, for \textit{Escherichia coli}, the cell-surface adhesion varied between $0.29\pm0.17$ nN to $0.39\pm0.20$ nN, showing a weak dependence on the substrate stiffness, thus suggesting that the stiffness-modulated adhesion is a species-specific trait. Finally, by quantifying the adhesion of \textit{C. okenii} population across different timescales, we report an emergent co-existence of weak and strongly adherent sub-populations, demonstrating a diversification of adherent phenotypes over the growth stages. Taken together, these findings suggest that bacteria, depending on the species and their physiological stage, may actively modulate cell-to-surface adhesion in response to the stiffness of soft surfaces. Our results suggest how bacteria could leverage stiffness-dependent adhesion, and the diversity therein, as functional traits to modulate initial attachment, colonization and proliferation on soft substrates during the early stages of biofilm development.

\end{sciabstract}


\section*{Introduction}

From living tissues to biomedical scaffolds, bacteria attach and colonize a wide range of surfaces, spanning orders of magnitude of stiffness \cite{Guimaraes2020, Serrano-Aroca2022}. Once attached, subsequent growth and accumulation the surface-associated bacterial populations are underpinned by mechanotransduction, ultimately leading to well-developed biofilm structures\cite{Dufrene2020, Maier2021, wang_bacterial_2023, Wittmann2023}. Surface sensing and mechano-response during initial stages of biofilm development \cite{Otto2002, Ellison2017, Hug2017, berne2018bacterial} are crucial for the long term fate of biofilms. At the scale of individual cells, the surface energy of underlying substrates, i.e., their hydrophobicity and hydrophilicity, influence bacterial adhesion \cite{Vadillo-Rodriguezh2005, Bayoudh2006, Yuan2017, kandemir_mechanical_2018, Yang2022}. In addition, van der Waals and electrostatic forces could regulate the cell-surface interactions \cite{Carniello2018, zhang2019}, thereby influencing biofilm growth. Recent results have also accounted for local osmotic pressure and poroelastic attributes of substrates in controlling biofilm growth and morphology \cite{asp2022spreading}.

Over the recent years, multiple studies have indicated at the possible role of the substrate stiffness on the bacteria-surface adhesion. Majority of these studies have been conducted on agarose and poly-ethylene glycol (PEG) hydrogels \cite{wang_bacterial_2023, guegan2014alteration, kandemir_mechanical_2018}, and on polymeric substrates including poly-dimethylsiloxane (PDMS) \cite{song2014stiffness, straub_bacterial_2019, bawazir_effect_2023} and composite thin films \cite{lichter2008, Wilms2020}. While different bacterial species have been covered in these studies, a generalized framework -- capturing the impact of stiffness on bacterial adhesion -- is yet to emerge, in part due to observable inconsistencies across the reported trends \cite{song2014stiffness, kolewe_fewer_2015,kolewe_bacterial_2018}. Traditionally, agarose, a linear polymer composed of alternating D-galactose and 3,6-anhydro-L-galactopyranose monomers\cite{lahaye_chemical_1991} has been frequently used in studying bacterial growth and biofilm assays. Agarose possesses high gel strength, transparency,\cite{bertasa2020agar} and non-toxicity, making them suitable for investigating bacterial adhesion, growth and self-organization on such surfaces \cite{wang_bacterial_2023, guegan2014alteration, You2018, you2019mono}.
More recently, low-melting-point (LMP) agarose has gained attention as a special form of agarose \cite{nandhakumar_evaluation_2011,guarrotxena_ag-nanoparticle_2012,rodriguez2020agarose} wherein the gelling temperature is decreased due to chemical modifications, for example via hydroxyethylation\cite{sambrook1989molecular}. The ability to form thermoreversible gels at low temperatures makes LMP agarose a preferred choice where gentle gelation conditions are desired, for instance, for bacterial assays \cite{Kim2018, Lewis2022,Wittmann2023}, encapsulation of heat-sensitive biomolecules\cite{raghunath2007biomaterials}, tissue engineering scaffolds\cite{raghunath2007biomaterials}, drug delivery systems\cite{dong2021smart,yazdi2020agarose}, and 3D cell culture models\cite{torabi_cassiebaxter_2019}. 

Despite the growing relevance of LMP agarose for bacterial studies, a thorough quantification of their mechanical properties are largely lacking. Bulk and local elasticity, stiffness and adhesion of substrates are of particular interest in microbial ecology, as they not only influence the physiology and behavior of cells, but also impact the biophysical underpinnings of bacterial growth, feedback and emergent traits \cite{Sengupta2020, Jin2024, Rani2024}. Elasticity is a fundamental mechanical property that characterizes the ability of a material to deform under an applied force and return to its original shape upon force removal. Various techniques, including bulk rheology and local atomic force microscopy (AFM) have been employed to investigate the elastic properties of such soft gels to understand their behavior under different loading conditions\cite{roberts_comparative_2011,normand_new_2000,watase_rheological_1983}. Adhesion, the ability of a material to stick to another material, may originate either due to mechanical interlocking or various physico-chemical interactions between surface molecules \cite{kreve_bacterial_2021}. In the context of bacterial adhesion, diverse species- and trait-specific attachment mechanisms have been proposed, which ultimately underpin the growth and dynamic self-organization into bacterial colonies\cite{araujo2023,you2019mono,genova2019mechanical,wang_bacterial_2023, dhar2022}. 
While recent studies have indicated that bacterial attachment and accumulation could be impacted by the stiffness of the underlying substrates, currently we lack methodologies which could allow simultaneous measurement of both parameters. Furthermore, how stiffness-dependent adhesion evolves with the age of a bacterial population remains unexplored. As described in the aforementioned studies, the impact of stiffness on cell-surface adhesion has received qualitative treatment so far, either by estimation of the surface density of the adhering cells, or via retention-assays which estimate the relative fraction of leftover cells following a washing protocol. Quantification of the cell-surface adhesion forces, alongside concomitant measurement of the underlying substrate stiffness is currently lacking. 

Motivated by the gaps in our current understanding, here we investigate the mechanical properties of LMP agarose gels prepared with different LMP agarose concentrations in Lysogeny broth (LB) medium, a nutrient-rich liquid used for growing bacteria. We measure their elasticity by Young's modulus, and the adhesion property, both native and with respect to different bacterial species, by means of cellular Force Distance Spectroscopy (FDS) technique based on an Atomic Force Microscopy (AFM), under the LB medium as well as deionized (DI) water. Focusing on the Gram negative bacterial species, \textit{Chromatium okenii}, we measure their cell-surface adhesion to the LMP agarose and analyze how their adhesion develops over the physiological growth stages. We compare our results with another Gram negative species, \textit{Escherichia coli}, as well as assess the possibility of active modulation of cell adhesion by contrasting the bacterial experiments against adhesion of passive polystyrene beads of 5$~\mu$m and 20$~\mu$m diameters to LMP agarose substrates. Overall, our findings reveal that that bacteria, depending on the species and their physiological state, can actively modulate cell-to-surface adhesion, as a response to changes in the stiffness of underlying soft surfaces. In doing so, bacteria may harness adherent phenotypes within a population to initiate attachment and optimize proliferation strategies on soft substrates across a range of stiffness. 

\section{Experimental}

\subsection{Force Distance Spectroscopy of bacterial cells on soft surfaces}
By means of a Nanoscope® V controller (Veeco) AFM with Multimode 8 scanner, force distance curves on pure agarose gels were measured using probes with a spherical $\ch{SiO2}$ tip (3.5~µm diameter, 0.2~N/m force constant, CP-CONT-SiO, nanoandmore GmbH, Germany). For cell adhesion measurement, tipless cantilever (0.2~N/m force constant, TL-CONT, nanoandmore GmbH, Germany) were used. Sample measurement was fulfilled in fluid (deionised (DI) water or LB medium) using a fluid chamber with inlets and rubber sealing. Deflection sensitivity was determined beforehand from the slope of the retract curve of a force distance spectroscopy measurement performed in measurement fluid on a silicon surface. To translate deflection to forces, the force constant measured by nanoandmore GmbH was used.
After removing the cover slip, the prepared sample holder was put in the AFM and a droplet of the measurement fluid was transferred onto the agarose gel surface. The prepared fluid chamber was installed and filled with measurement fluid through the inlets using syringes. Subsequently, the cantilever was manually approached to the surface but not brought to contact. During this procedure, the horizontal position of the cantilever was moved to the center position of the region of interest. 
After visually close tip and surface approximation, the probe was false engaged (i.e. tip-surface contact was simulated) and the sample was subsequently manually approached by means of the step motor control and continuously measured force distance curves until an increase in slope was detectable.
For cell and microparticle adhesion measurement, the cantilever tip region was first approached towards cell accumulation regions. A cell/microparticle was picked up by the tipless cantilever and used as cantilever tip. The subsequent measurements were executed at a cell/microparticle-free region on the sample surface.
Force spectroscopy curves were recorded on 16~×~16 equidistant spots around the initial tip-surface contact point. The distance between adjacent spots was 500~nm. The measurements were repeated at least 4 times on the same sample. Maximum applied force did not exceed 10 nN. 

\subsection{Optical Microscopy}
To ensure that cells and microparticles were present on the tipless cantilever, the sample holder as well as the tip holder were both investigated under an optical microscope (Eclipse LV100N POL, Nikon Corporation) in reflected light mode right after measurement and careful removal of excess liquid with a soft absorbant tissue.

\subsection{Evalution of the Force Distance Spectroscopy data}
AtomicJ (v 2.3.1, Pawel Hermanowicz) was used to extract the measured values from the recorded data files and process them by applying different theoretical contact mechanic models. A typical force distance curve is shown schematically in Figure~\ref{fig1:ExperimentalSetup}A. It consists of a region of no contact between tip and substrate and a region of indentation. These two regions are connected by the Contact Point $z_c$. For proper calculation, precise determination of the position of the contact point is indispensable.
Contact mechanic models are applied to the contact area of the force distance curves. They are based on the Hertz model\cite{hertz1882ueber}, which describes the interaction between two spheres or a sphere and an endless, plane surface (see Figure \ref{fig1:ExperimentalSetup}B). According to Hertz, the generalised force-indentation relation can be described by:

\begin{equation}
    F=\lambda \delta ^\beta
\end{equation}

where $F$ is the applied force, $\delta$ is the indentation depth, $\lambda$ and $\beta$ are values that depend on the underlying model. For the basic Hertz model indentation of a spherical body into a plane surface, $\beta$ is 3/2 and $\lambda $ is given by the following equation:

\begin{equation}
    F = \frac{4ER^{1/2}}{3(1-\upsilon^2)}\delta^{3/2}
\label{eq:YM_Hertz}
\end{equation}

Here, $E$ is the elasticity modulus, $R$ is the tip radius and $\upsilon$ is the Poisson Ratio. The latter is assumed to be $\approx$~0.5 for very soft samples.

The Hertz model is a very basic model that does not consider adhesion forces and their subsequent contact area enhancement. Derjaguin, Muller and Toporov (DMT)\cite{derjaguin1975effect} as well as Johnson, Kendall and Roberts, JKR \cite{johnson1971surface}, have parallely created extended models based on the Hertz model to incorporate adhesion forces. These two models represent the limit values for the true value of elasticity. While DMT is more accurate for harder samples with low adhesion, soft samples with higher adhesion are better described by JKR.

In this paper we therefore focus on the evaluation using the JKR model via the following three equations:

\begin{equation}
\delta = \frac{\alpha_{\text{JKR}}^2}{R} -\frac{4}{3} \sqrt{\frac{\alpha_{\text{JKR}} F_{\text{ad}}}{R K}}  
\label{eq:YM_JKR1}
\end{equation}
\begin{equation}
\alpha_{\text{JKR}} = \left[ \frac{R}{K} \left( \sqrt{F_{\text{ad}}} + \sqrt{F + F_{\text{ad}}} \right)^{2}\right]^\frac{1}{3}
\label{eq:YM_JKR2}
\end{equation}
\begin{equation}
F_{\text{ad}} = \frac{3}{2} \pi \gamma R
\label{eq:YM_JKR3}
\end{equation}
\begin{equation}
K = \frac{4E}{3(1-\upsilon^2)}
\label{eq:YM_JKR4}
\end{equation}

where $ \alpha_{\text{JKR}} $ is the JKR-extended contact area between tip and sample, $ F_{\text{ad}} $ is the tip-sample adhesion force, $K$ is the elastic constant of the sample, and $\gamma$ is the interfacial energy. The adhesion force, $F_{ad}$, was evaluated from the minimum point of the force distance curve relative to its baseline. 

$E$ can hence be calculated by plotting $\delta$ versus $F$ and adjust $K$ to minimize the least square error between the resultant plot and the measured data, starting from the tip-sample contact point. This latter is  ascertained numerically for best fit. More comprehensive details on force spectroscopy data evolution is well described by the works of Lin, Dimitriadis, and Horkay\cite{lin_robust_2007,lin_robust_2007-1}. The evaluated data are represented as boxplots without whiskers. The inner 50~$\%$ of the measured values sorted in ascending order are represented by a coloured area containing the median as a horizontal line. Outliers were identified as data points that lie below $Q_1 - 1.5 * IQR$ or above $Q_3 + 1.5 * IQR$, where $Q_1$ and $Q_3$ are the first and third quartiles, respectively, and $IQR$ is the interquartile range ($Q_3 - Q_1$). Outliers are not shown here. In addition, the average value is shown as a point and the standard deviation as an error bar.

\subsection{Low-melting-point agarose gel preparation}
Agarose gels with different concentrations (see Table~\ref{tbl1:AgaroseGelConcentrations}) were prepared by slowly dissolving the respective quantity of low-melting agarose gel (Agarose, LMP, Preparative Grade for Large Fragments ($>$ 1,000 bp), Promega, USA) in 5~ ml LB medium in an autoclaved glass tube under occasional stirring on a heat plate (50~\textdegree C). When the powder was completely dissolved, the solution was left to cool down to room temperature and then stored in a fridge (4~\textdegree C).

\begin{table}[H]
\begin{center}
  \caption{Agarose concentrations}
  \label{tbl1:AgaroseGelConcentrations}
  \begin{tabular}{ccc}
    \hline
    \centering  [g/ml] & [\%] \cr
    \hline
    0.0150 & 1.5\cr
    0.0183 & 1.8\cr
    0.0221 & 2.2\cr
    0.0262 & 2.6\cr
    0.0306 & 3.1\cr
    \hline
  \end{tabular}
  \end{center}
\end{table}

\begin{figure}[]
  \centering
  \includegraphics[width=0.9\textwidth]{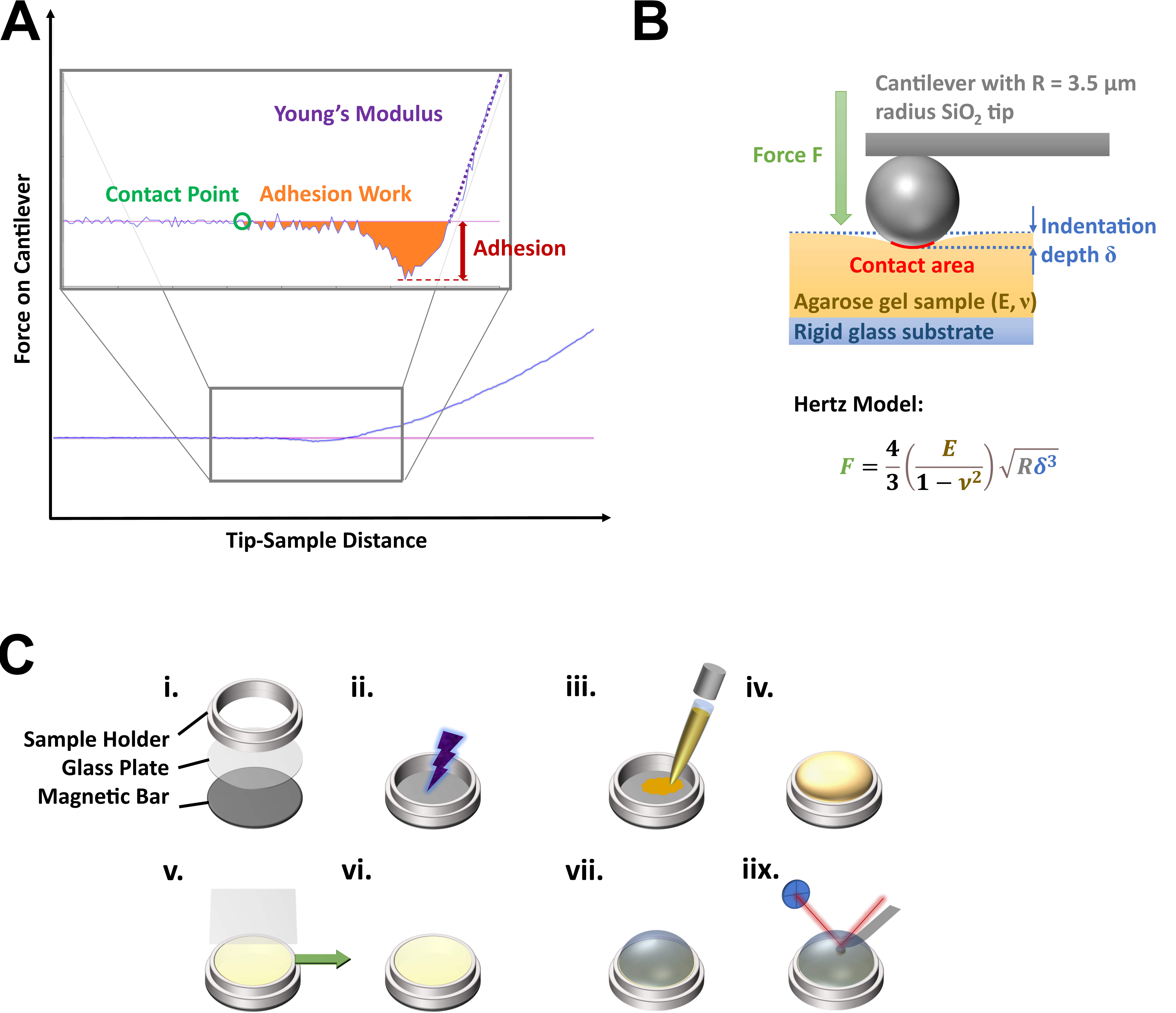}
  \caption{Experimental evaluation and setup. A) Example Force Distance Curve. The inlet shows an enlargement of the dashed area. Adhesion force is extracted from the lowest force relative to the baseline. Young's modulus is calculated from the slope of the curve. Maximum indentation force was 100 nN. B) Hertz Model when a solid sphere with radius~$R$ indents a sample with Young's modulus~$E$ and Poisson Ratio~$v$. C) Sample preparation: The sample holder consists of a magnetic bar to immobilize the holder on the AFM scanner, a glass plate as bottom and the sample holder confinement (i). The glass is plasma-treated (ii) for 45~s. Agarose gel is heated up to liquefaction and then transferred to the sample holder (iii) until complete filling (iv). While still liquid, a glass coverslip is placed onto the sample holder while avoiding air entrapment (v). The sample is allowed to gelify and cool down for 30~min. Before measurement, the coverslip is removed horizontally (v, vi). The sample is placed onto the AFM and the measurement liquid is injected (vii). After 15~min of relaxation time, the measurement begins (ix).
}
  \label{fig1:ExperimentalSetup}
\end{figure}

Sample preparation is shown in Figure~\ref{fig1:ExperimentalSetup}C. A ring-shaped sample holder with an inner diameter of 10~mm and a height of 2~mm was glued onto a glass coverslip with a diameter of 12~mm so that no glue was inside the ring. A magnetic plate was glued to the other side of the coverslip. The whole setup was dried for 24~hours and used as AFM sample holder.
Prior to sample preparation, the agarose gel was taken out of the fridge and slowly heated on a heater plate. In order to increase hydrophilicity of the sample holder glass, the sample holder glass surface was plasma-treated for 45 seconds using a plasma torch (Mode BD-20, Electro-Technic Products, Inc., USA). 200~µl of the warm and viscous agarose gel was transferred to the sample holder. A coverslip was placed onto the sample holder so as to enclose the agarose gel while avoiding to trap air inside. The agarose gel was left to cool down for 30~minutes, then the coverslip was slowly and horizontally removed just prior to measurement. 
The topographies of two samples with 2.2~\% agarose concentration were measured (in LB medium environment and in water environment) by means of a Nanoscope® V controller (Veeco) AFM with Multimode 8 scanner in Tapping mode using probes with a spherical $\ch{SiO2}$ tip (3.5~µm diameter, 0.2~N/m force constant, CP-CONT-SiO, nanoandmore GmbH, Germany).

\newpage

\subsection{Adhesion of polystyrene beads to LMP agarose gel}
Polystyrene microparticles with 5~µm and 20~µm were purchased from microparticles GmbH, Germany (product numbers PS/Q-R-KM650 and PS/Q-R-KM636 respectively). The stock solutions were diluted by factor $1 \times 10^{-3}$ until single particles were visible on the substrate after sedimentation. The agarose gel sample was prepared as described above. After removement of the coverslip from the agarose gel sample, a drop of the microparticles solution was put onto the agarose gel sample. The microparticles were allowed to settle down for 30 minutes before the drop was carefully removed. Subsequently, force distance curves were recorded as described above. This procedure was repeated three times. In order to measure the adhesion of the passive particles, the same measurements were repeated with agarose gel based on DI water as well. For each agarose concentration and particle size, the procedure was repeated three times.

\subsection{Preparation of bacterial assays}
Adhesion forces of two different cell species \textit{Escherichia coli} (strain NCM 3722 $\Delta$motA, referred to as \textit{E. coli}) and \textit{Chromatium okenii} (referred to as \textit{C. okenii}) to LMP agarose gel with different concentrations (1.5~\%, 2.2~\% and 3.1~\%) were determined.

 \textit{E. coli} were stored in a freezer at -80~\textdegree C. Agar plates were prepared by dissolving 16~g of LB-agar (LB-Agar (Luria/Miller), CARL ROTH, Germany) in 400~ml deionized water with subsequent autoclaving and stored at 4~\textdegree C. The cells were taken out of the freezer and streaked on an agar plate in sterile environment at room temperature. The agar plate was placed in an incubator for 48~h at 30~\textdegree C. Subsequently, a single cell island was transferred from the agar plate into 8~ml LB medium using a bacterial loop in sterile environment. The cell solution was placed into the incubator at 30~\textdegree C and shaken at 180~RPM during 24~hours. Thereafter, 200~µl of this cell solution was transferred to 2~ml LB medium and kept in the incubator at 30~\textdegree C under continuous shaking at 180~RPM. After 160~min, the cell suspension was centrifuged at 2000~rpm for 5~min. Except for 100~µl, the centrifugate was slowly removed. The cells were resuspended and transferred onto a freshly prepared agarose gel surface sample (described above). The cells were allowed to settle down on the agarose gel surface for 30~minutes. The drop was then carefully removed and FDS was executed as described below, under the LB medium environment.

Phototrophic sulfur bacteria \textit{C. okenii} was obtained from Lake Cadagno in the Piora Valley (46°33’N, 8°43’E) in the southern Swiss Alps, following the protocol described in Ref. \cite{Sommer2017}. They play a key role in sulfide removal from shallow sediments and stratified waters \cite{Berg2019}. Upon transferring the cells to laboratory, they were grown and propagated in Pfennig's Medium I protocol with periodic spiking of hydrogen sulphide\cite{Nezio-front2023}. Cultivation was run under light/dark photoperiod (16/8~h) with a light intensity of 38.9~µmol $m^{-2}$ $s^{-1}$. After soft shaking, 100~µl of the cell solution was transferred onto the freshly prepared LMP agarose surface (described above). The cells were allowed to settle down for 30~minutes. Subsequently, the drop was carefully removed, and measurement was executed as described below under acqueous environment. 

\subsection{Quantification of \textit{C. okenii} adhesion properties over growth stages}

In order to track the evolution of cell-surface adhesion over time, FDS was carried after 2~weeks, 10~weeks and 14~weeks of inoculation of \textit{C. okenii} in Pfennig's Medium I. Although motile in their natural lake ecosystem, under laboratory conditions, \textit{C. okenii} gradually shifts to a sessile lifeform \cite{di2023synergistic}. The physiology-dependent measurements were conducted on agarose concentration of 2.2~\%, following the gel preparation protocol described previously. The cell solution was kept in a glass container enclosed with a septum so as to protect it from air contact. For each measurement, 0.8 ml of cell solution was taken using a 1 ml syringe and needle. The cells were collected either from the surface of the cell solution (top cells) or from the bottom of the glass container (bottom cells) and transferred to a 2 ml Eppendorf tube. After removal, the cells were centrifuged at 6000~RPM for 180~s. The pellet was carefully aspirated using a 10 µl pipette and finally transferred to the prepared LMP agarose gel substrate. The sample was then left to stand for 30 minutes so that the cells could settle on the substrate. The cantilever was wetted with a drop of DI water before the cantilever holder was installed. This gentle method minimised fluid flow which could interfere measurements within the sample solution. Each measurement series consisted of six measurements. Firstly, three measurements were carried out with top cells. Then three measurements were taken with bottom cells. Each series of measurements was completed within 3 days.

\section{Results}

\subsection{Topography of LMP agarose gel surfaces}
\begin{figure}[H]
  \centering
  \includegraphics[width=0.9\textwidth]{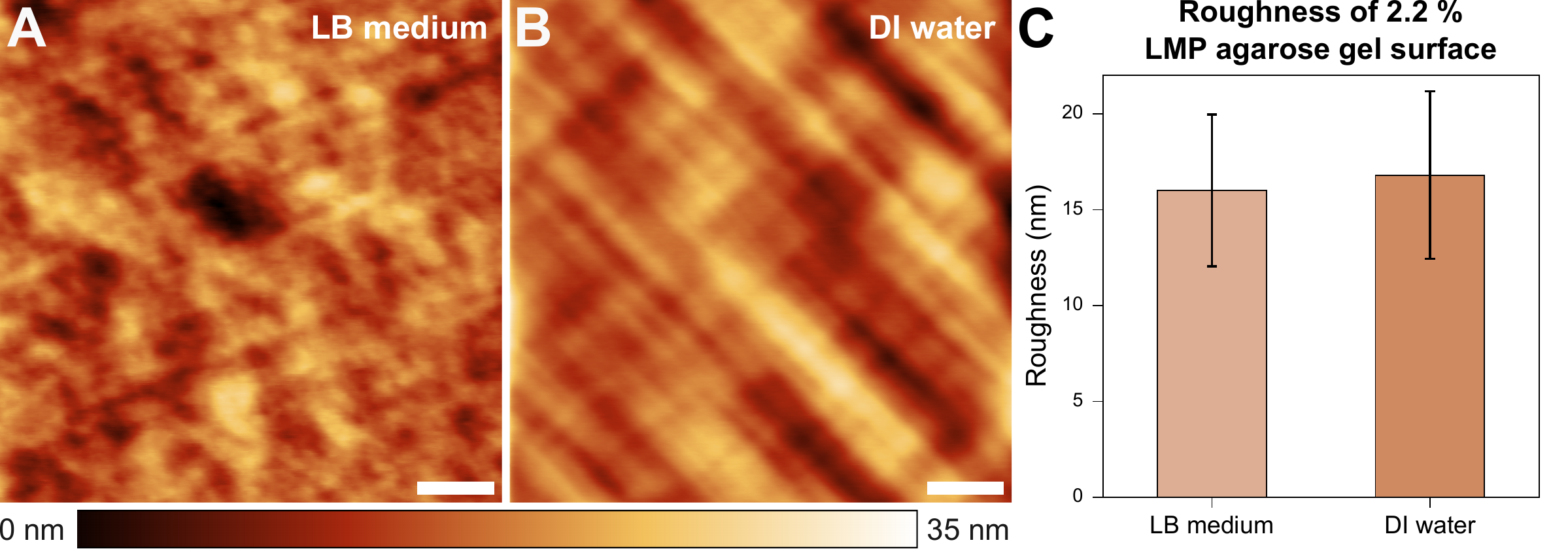}
  \caption{Sample topographies on a 15~x~15~µm² area of 2.2~\% LMP agarose gel concentration samples in LB medium (A) and water (B) environment. The roughness is 16.01~($\pm$ 3.96)~nm, and 16.80~($\pm$ 4.37)~nm, respectively (C). The scale bars indicate 3~$\mu$m.
}
  \label{fig2:Topography}
\end{figure}

Figure~\ref{fig2:Topography} shows the topography of 2.2~\% agarose gel in LB medium (A) and DI water (B) environments. The sample surfaces showed overall low surface roughness, furthermore, no significant difference in roughness was observed between the two cases. 

\subsection{Mechanical properties of the LMP agarose gels}
\begin{figure}[H]
  \centering
  \includegraphics[width=\linewidth]{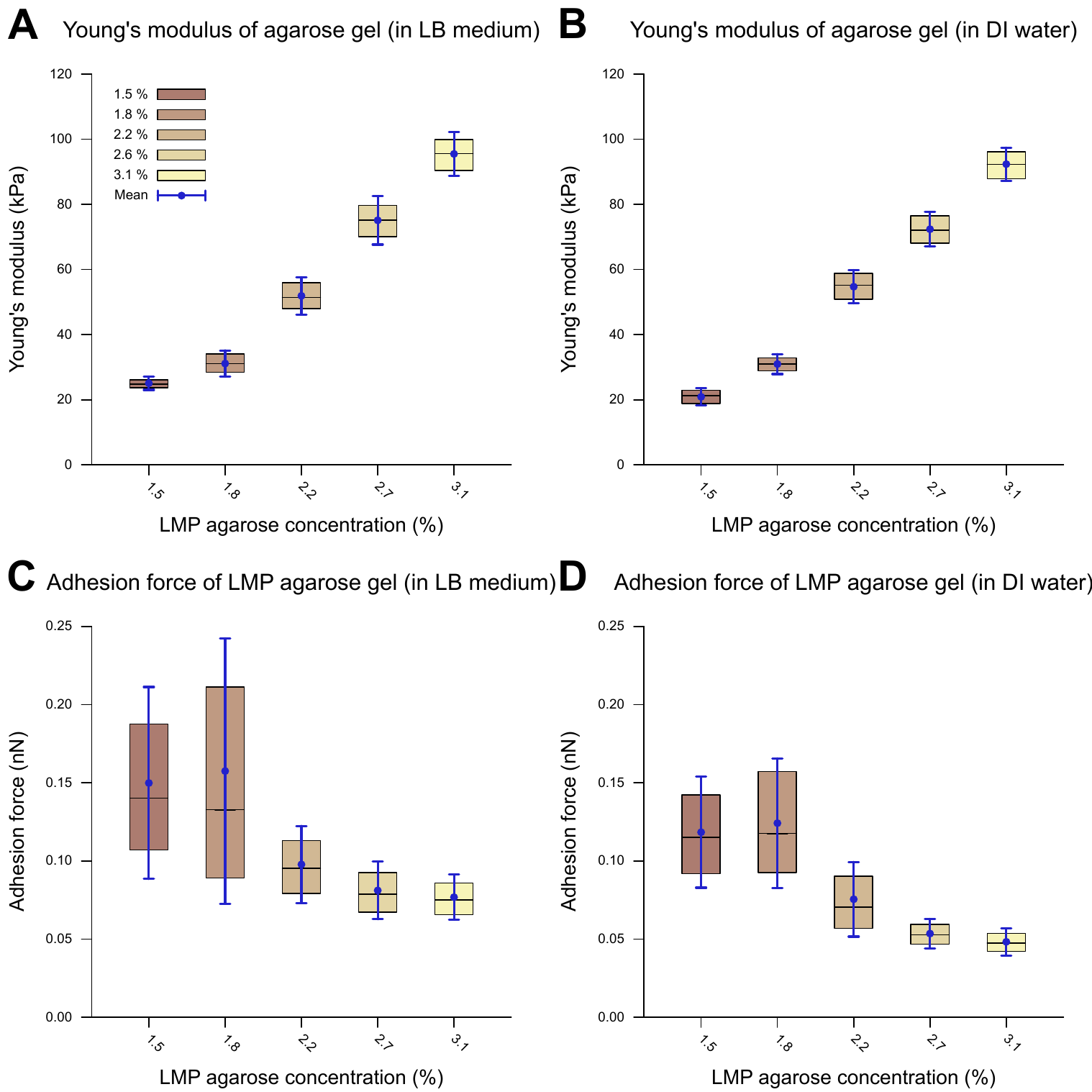}
  \caption{Measured mechanical properties of agarose gel with different concentrations. The inner 50~$\%$ of the measured values sorted in ascending order are represented by a coloured area containing the median as a horizontal line. In addition, the average is shown as a point and the standard deviation as an error bar. The difference between water and LB medium as measurement environment is imperceptible. The Young’s modulus of agarose gel increases with rising LMP agarose concentration and lies between 20 and 100~kPa in both LB medium (A) and water (B) environment. Only small adhesion forces in sub-nN scale could be observed for both LB medium (C) and water (D) environment in contact to a spherical \ch{SiO2} tip (3.5~µm). Adhesion forces decrease slightly with increasing LMP agarose concentration.
}
  \label{fig3:Stiffness}
\end{figure}

A representative example for a measured force distance curve is given in Figure~\ref{fig1:ExperimentalSetup}A. The computed outcomes of Young's modulus and adhesion values of agarose gel samples without cells are presented in Table~\ref{tbl2:Stiffness}. Figure~\ref{fig3:Stiffness} shows the corresponding measurements conducted in DI water and LB medium. The provided data includes both median and mean values ($n > 500$).

\begin{table}[H]
\centering
  \caption{Young's modulus of agarose gel samples with different agarose concentrations}
  \label{tbl2:Stiffness}
  \begin{tabular}{c:c:cc:cc:c}
    \hline & Concentration & \multicolumn{2}{c:}{Young's modulus} & \multicolumn{2}{c:}{Adhesion Force} & Number of\\

    &  & Average & Median & Average & Median & measurements\\
     & [\%] & [kPa] & [kPa] & [nN] & [nN] & (\textit{n})\\
    \hline
    \multirow{5}{*}{\rotatebox[origin=c]{90}{LB Medium}} & 1.5 & 25.1~($\pm$~2.1) & 24.8 & 0.15~($\pm$~0.06) & 0.14 & 881 \\
    & 1.8 & 31.1~($\pm$~4.0) & 31.1 & 0.16~($\pm$~0.08) & 0.13 & 1083 \\
    & 2.2 & 51.9~($\pm$~5.7) & 51.4 & 0.10~($\pm$~0.02) & 0.10 & 1069 \\
    & 2.6 & 75.1~($\pm$~7.5) & 75.2 & 0.08~($\pm$~0.02) & 0.08 & 589 \\
    & 3.1 & 95.5~($\pm$~6.7) & 95.6 & 0.08~($\pm$~0.01) & 0.07 & 529 \\
    \hline
    \multirow{5}{*}{\rotatebox[origin=c]{90}{Water}} & 1.5 & 20.9~($\pm$~2.6) & 21.2 & 0.12~($\pm$~0.03) & 0.11 & 557 \\
    & 1.8 & 31.0~($\pm$~3.1) & 31.0 & 0.13~($\pm$~0.04) & 0.12 & 589 \\
    & 2.2 & 54.7~($\pm$~5.1) & 55.2 & 0.08~($\pm$~0.02) & 0.07 & 931 \\
    & 2.6 & 72.4~($\pm$~5.3) & 72.0 & 0.05~($\pm$~0.01) & 0.05 & 522 \\
    & 3.1 & 92.4~($\pm$~5.1) & 92.3 & 0.05~($\pm$~0.01) & 0.05 & 503 \\
    \hline
    \end{tabular}

\end{table}

The Young's modulus, also known as the modulus of elasticity, is a measure of the stiffness of a material, defined as the ratio of stress to strain within the elastic limit of the material. Upon increasing the LMP agarose concentration, the Young's modulus increases, for both DI water and LB medium environments. With Young's moduli smaller than 100 kPa, the agarose substrate falls within the range of soft biomaterials (e.g., cells, organs) and soft synthetic polymers\cite{cox2011remodeling,sachot2014hybrid}. The observed trend can be elucidated by examining the underlying structure of agarose gels\cite{arnott1974agarose,lahaye_chemical_1991,md21050299}. LMP agarose monomers consist $\beta$-D-galactose and 3,6-anhydro-$\alpha$-L-galactose units. Upon cooling the agarose gel solution below the gelation temperature (24 - 28~\textdegree C), the equatorial hydrogen atoms of the 3,6-anhydro-$\alpha$-L-galactose residues form hydrogen bonds with each other, inducing the formation of $\alpha$-helices, forming a single chain or double chains. The cooling rate determines the ratio of one-chain and two-chains $\alpha$-helices. This process results in the formation of a 3D secondary agarose gel structure. The gelation temperature and pore size of the gel primarily depend on the availability of hydrogen and oxygen in the agarose solution. A lower agarose concentration leads to fewer $\alpha$-helices, resulting in reduced structural stability and larger pore sizes. Alongside $\alpha$-helices, other types of physical cross-linking between the chains increase the stability of the gel matrix. The structural stability directly correlates with the gel's Young's modulus, explaining the increase in Young's modulus with higher agarose concentration. Low-melting agarose is produced by hydroxyethylation such that a portion of the available alcoholic side groups is substituted by ethyl groups and hence unable to form hydrogen bonds. This substitution reduces the density and strength of $\alpha$-helices, leading to a lower Young's modulus\cite{serwer1983agarose}. Overall, the Young's moduli we measure are lower than those of normal melting agarose gels, as reported in literature\cite{normand_new_2000}. Roberts \textit{et al.} investigated the elastic storage modulus of 1~$\%$ normal agarose gel to be around 80~kPa\cite{roberts_comparative_2011}, with other reports corroborating their results \cite{nijenhuis1997thermoreversible,normand_new_2000,guenet2006agarose,fernandez_rheological_2008,mao2017dynamics}.

The mechanical properties of LMP agarose gels in LB medium have received limited investigation so far. Existing studies report varying results on mechanical properties across different agarose concentrations, though the overall range matches well with our measurements. Kontomaris \textit{et al.} used a similar LMP agarose gel and measured an elastic modulus of 154~kPa (2.5~$\%$)\cite{kontomaris20233d}, while Topuz \textit{et al.} observed values in the same range as this study for LMP agarose gel\cite{topuz2018nanosilicate}. Finally, Zamora-Mora \textit{et al.} measured an Young's modulus of 400~Pa for 0.5~$\%$ agarose gel and 1.45~kPa for 2~$\%$ agarose concentration.\cite{zamora-mora_chitosanagarose_2014}, similar to Kumachev \textit{et al.}. who reported the range to be between 50~Pa (0.75~$\%$) to 2.5~kPa (3.00~$\%$) \cite{kumachev_high-throughput_2011}. It is to noted that last two studies measured the shear modulus, though the Young's modulus is expected to be in the same range as well.

\newpage

\subsection{Cell-surface adhesion force on LMP agarose gels}
\begin{figure}[H]
  \centering
  \includegraphics[width=\linewidth]{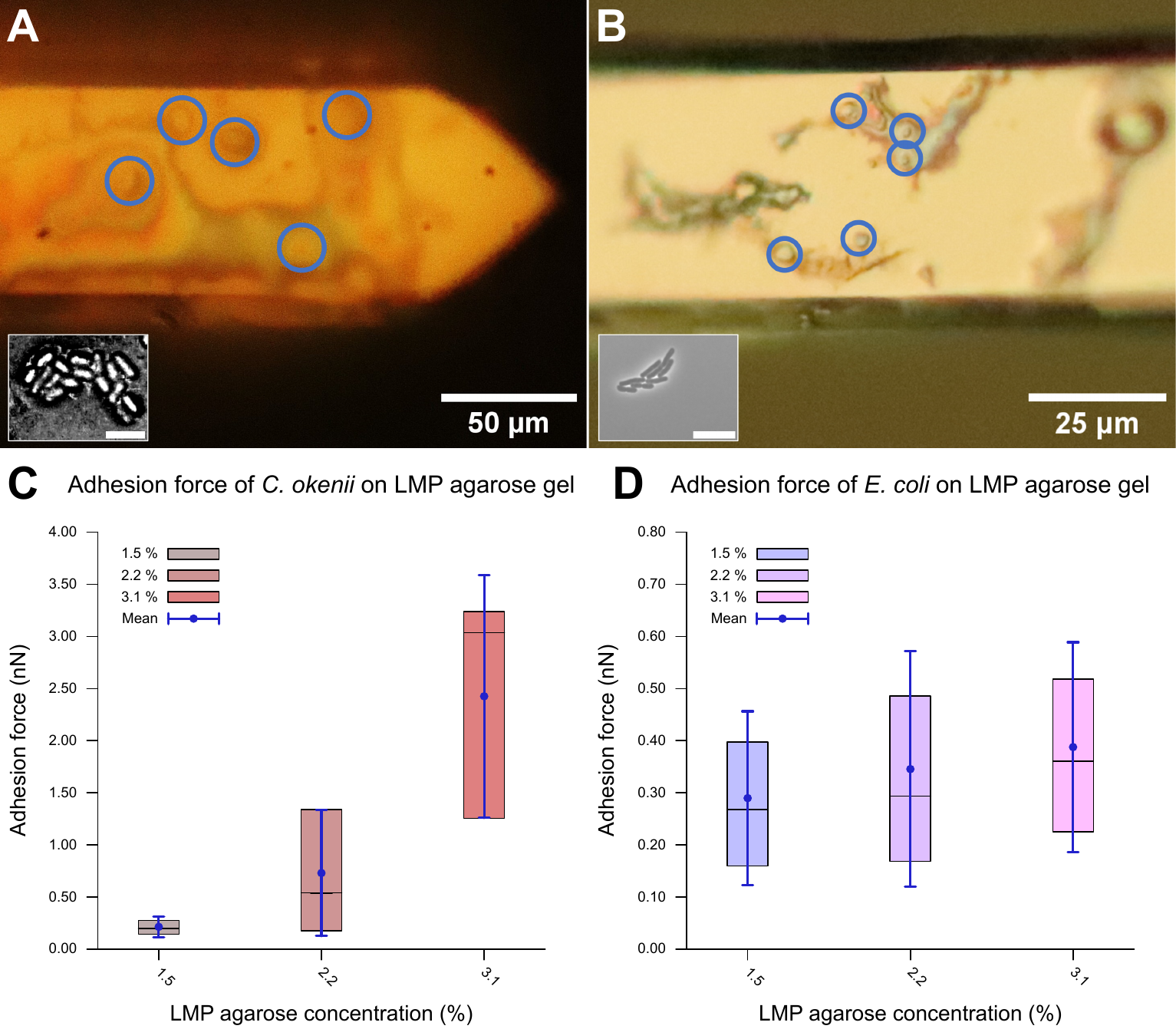}
  \caption{
  Adhesion properties of \textit{E. coli} and \textit{C. okenii} on agarose gel with varying concentration. By utilizing optical microscopy, the cells on the cantilevers were clearly observed, as shown in (A) for \textit{E. coli} and (B) for \textit{C. okenii}. The presence of cells on the cantilever (encircled) provided compelling evidence that the adhesion between the gel surface and the cells was successfully captured by the measurement. The insets display optical microscopy images of \textit{E. coli} (A) and \textit{C. okenii} (B) on agarose gel, with a scale bar of 10 µm. In C and D, the inner 50~$\%$ of the measured adhesion force values towards LMP agarose gel sorted in ascending order are represented by a coloured area containing the median as a horizontal line. In addition, the average is shown as a point and the standard deviation as an error bar Interestingly, as the LMP agarose concentration was increased, a slight rise in adhesion force for \textit{E. coli} bacterial cells (C) and a more substantial increase for \textit{C. okenii} bacterial cells (D) were observed.
}
  \label{fig4:AdhesionBacteria}
\end{figure}

The evaluated values for adhesion of \textit{E. coli} and \textit{C. okenii} cells to LMP agarose gel with different agarose concentrations are listed in Table~\ref{tbl3:AdhesionBacteria} and plotted in Figure~\ref{fig4:AdhesionBacteria}. Optical microscopy images (Figure~\ref{fig4:AdhesionBacteria}A and B) reveal that cells were attached to the cantilever surface, confirming that the adhesion force between cell and agarose gel surface was measured. The force distance curves show a smooth adhesion characteristic, as has been observed for cell initial adhesion on hydrophilic surfaces\cite{maikranz2020different}. For \textit{E. coli}, the adhesion values are in weak sub-nN range. As can be seen, adhesion forces slightly increase as agarose concentration increases, which is in agreement with the adhesion to normal agarose gels\cite{guegan2014alteration}.

\begin{table}[H]
\begin{center}
  \caption{Adhesion force of bacteria to agarose gel of different concentrations}
  \label{tbl3:AdhesionBacteria}
  \begin{tabular}{ccccc}
    \hline
    \centering Strain & Concentration [\%] & Average [nN] & Median [nN] & Measurements \cr
    \hline
\textit{E. coli} & 1.5 & 0.29~($\pm$~0.17) & 0.27 & 482\\
\textit{E. coli} & 2.2 & 0.35~($\pm$~0.23) & 0.29 & 722\\
\textit{E. coli} & 3.1 & 0.39~($\pm$~0.20) & 0.36 & 917\\
\hline
\textit{C. okenii} & 1.5 & 0.21~($\pm$~0.10) & 0.20 & 862\\
\textit{C. okenii} & 2.2 & 0.73~($\pm$~0.60) & 0.54 & 1476\\
\textit{C. okenii} & 3.1 & 2.42~($\pm$~1.16) & 3.04 & 443\\
    \hline
  \end{tabular}
\end{center}
\end{table}

\begin{figure}
  \centering
\includegraphics[width=0.47\linewidth]{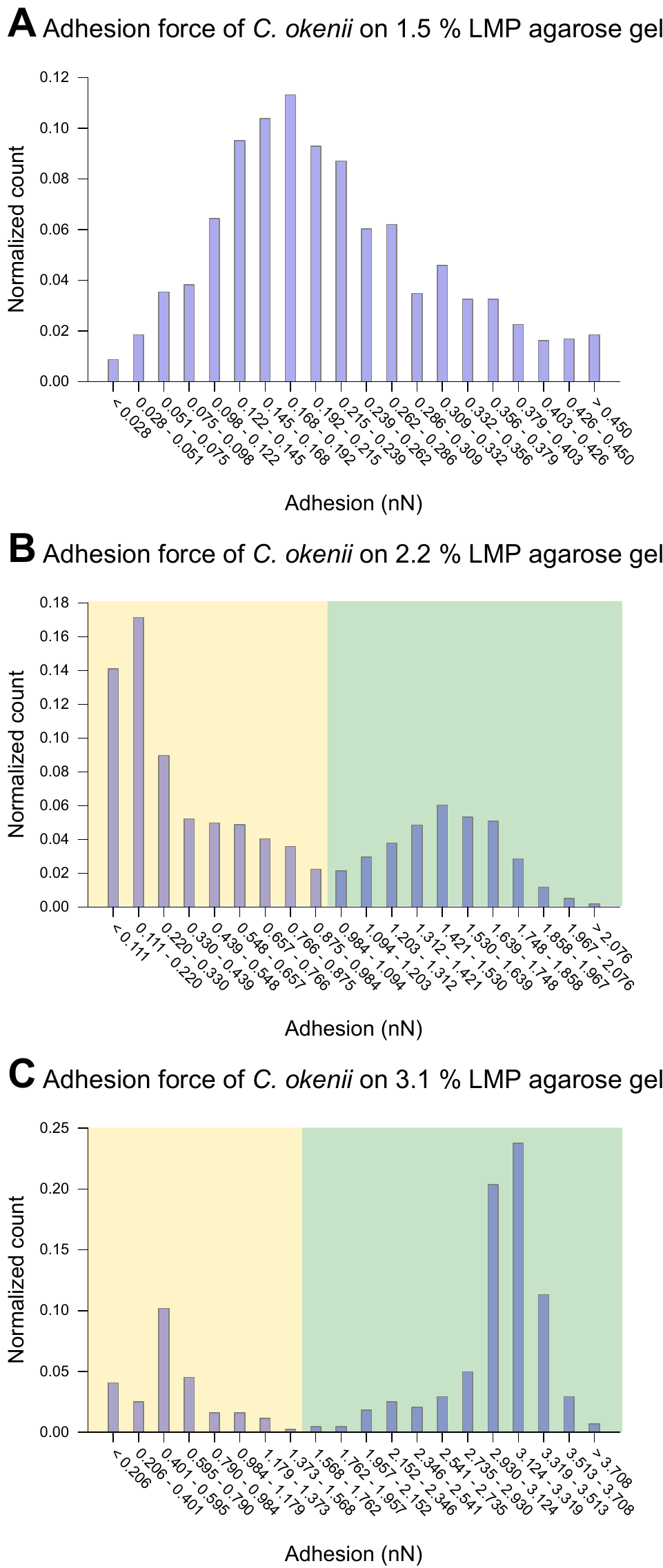}
  \caption{Adhesion force distribution of \textit{C. okenii} on 1.5~\% LMP agarose gel (A), 2.2~\% LMP agarose gel (B) and 3.1~\% LMP agarose gel (C). While for 1.5~\% LMP agarose gel the adhesion force is unimodal, two peaks occur for 2.2~\% (B) and 3.1~\% (C) agarose concentrations. A manually threshold was set to divide the occuring peaks into lower adhesion force part (highlighted in yellow) and higher adhesion part (highlighted in green).}
  \label{fig5:SingleAdhesionDistribution}
 
\end{figure}

For \textit{C. okenii}, however, the adhesion force increased by up to an order of magnitude compared to \textit{E. coli} adhesion forces. Furthermore, the \textit{C. okenii} are found to be more sensitive to changes in the LMP agarose concentration. While doubling the agarose concentration from 1.5~\% to 3.1~\% increases the median of adhesion force surface by factor~1.3 for \textit{E. coli} cells, it is enhances by a factor of ~15 for \textit{C. okenii}. Furthermore, a detailed examination of the adhesion force data distribution for \textit{C. okenii} reveals emergence of a bimodal distribution of adhesion force at LMP agarose concentrations of 2.2~\%, which then becomes more pronounced when the agarose concentration is raised to 3.1~\% (see Figure~\ref{fig5:SingleAdhesionDistribution}). In order to understand the co-existence of the adherent phenotypes, we separated the two peaks in the bimodal adhesion force distribution for 2.2~$\%$ and 3.1~$\%$ agarose concentrations, with dynamic threshold adhesion values, as indicated by the yellow and green regions in Figure~\ref{fig5:SingleAdhesionDistribution}. The bimodal distributions indicate the co-existence of two different adherent sub-populations: weakly adherent cells (yellow regions of the plot), and the strongly adherent cells (green regions of the plot), which emerge as the stiffness of the underlying substrate is modulated. At an LMP concentration of 1.5~$\%$, such a bimodality could not be detected.

\begin{figure}
  \centering
  \includegraphics[width=0.9\textwidth]{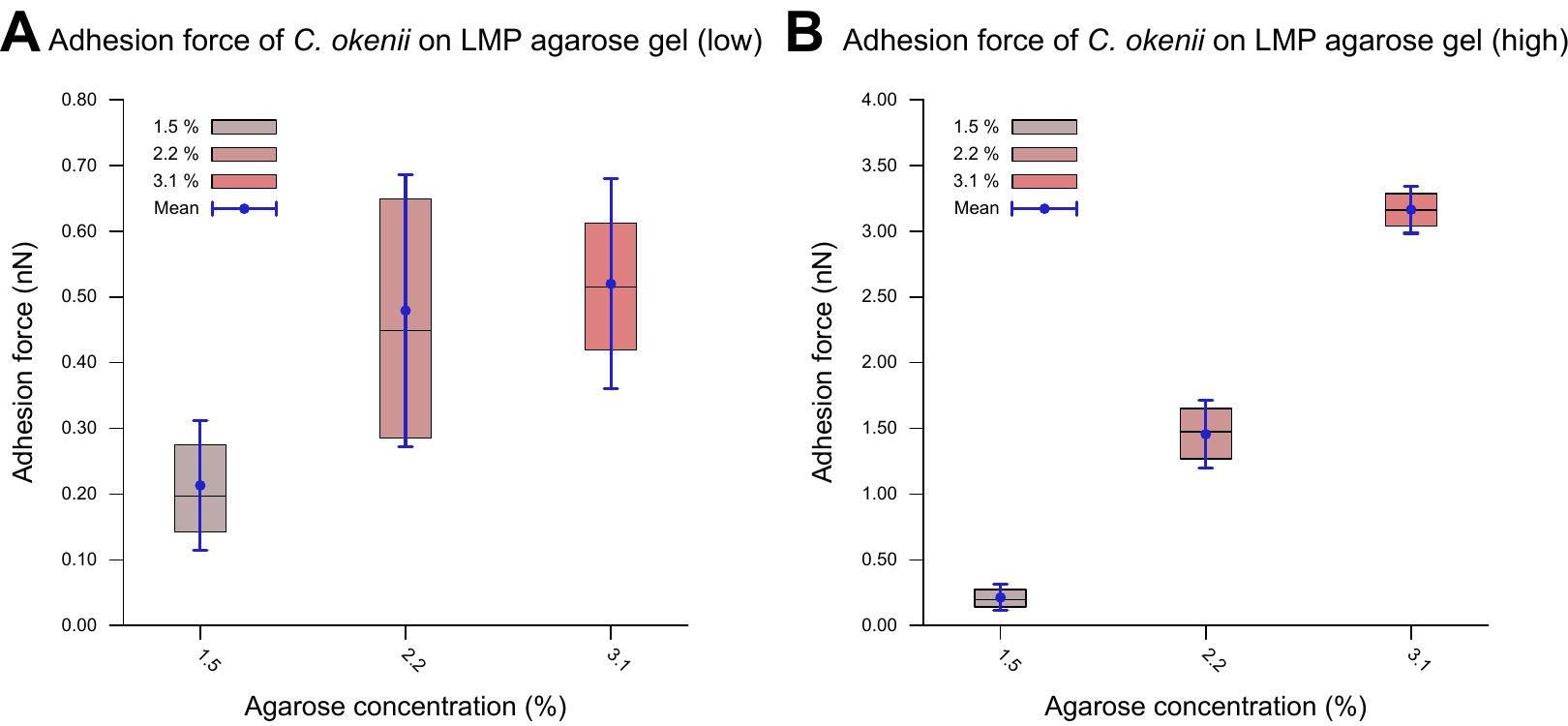}
  \caption{Adhesion force of \textit{E. coli} and \textit{C. okenii} towards LMP agarose gel with various concentrations. The inner 50~$\%$ of the measured values sorted in ascending order are represented by a coloured area containing the median as a horizontal line. In addition, the average is shown as a point and the standard deviation as an error bar. While \textit{E. coli} shows little adhesion (A), the adhesion force of \textit{C. okenii} strongly depends on LMP agarose concentration (B). Adhesion force values for 2.2~\% and 3.1~\% LMP agarose concentrations are bimodally distributed (see Figure~\ref{fig5:SingleAdhesionDistribution}). Seperating these peaks and replotting shows weak adhesion force for the lower adhesion part (C), comparable to (A), while the higher adhesion peaks show stronger adhesion forces that clearly depend on LMP agarose concentration (D). The occurrence of two adhesion force peaks signify two different adhesion mechanisms of \textit{C. okenii} towards LMP agarose gel.}
  \label{fig6:BoxplotsResolved}
\end{figure}

As shown in the boxplots, Figure~\ref{fig6:BoxplotsResolved}, the lower range of \textit{C. okenii} adhesion force distribution peaks around the same range as for \textit{E. coli} cells, i.e., low adhesion force that depends weakly on the agarose concentration. In contrast, the higher adhesion force peaks seen in \textit{C. okenii}, reveal a stronger dependence on LMP agarose concentration. The small adhesion noted in both \textit{E. coli} and \textit{C. okenii} corresponds with the general observation that bacteria with hydrophobic surfaces tend to adhere more readily to hydrophobic than hydrophilic surfaces. Nevertheless, the weak adhesion forces observed here, vary slightly as the agarose concentration is increased. To understand this, a bacteria can be modeled as solid body, with biopolymers on their surface represented by flexible chains\cite{maikranz2020different}. During the separation of a cell from the surface, various forces including electrostatic forces, hydrogen bonds, dipole interactions, and Lifshitz-Van der Waals forces act to oppose detachment. Taken together, the interactions can be captured using the extended DLVO theory (xDLVO)\cite{van1994interfacial}, however, in a more complex biological context (as the one studied here), the xDLVO theory might require further modifications, which are beyond the scope of this work. Another possible effect during the detachment of a bacterial cell is steric hindrance, due to the partial entanglement of the polymeric surface chains \cite{maikranz2020different}. The minimum force required for complete separation corresponds to the adhesion force, which depends on the properties of the flexible surface chains and the structure and properties of the underlying network, which in turn, alter with agarose concentration, offering a possible reason for the slight enhancement of the weak adhesion forces with increase in the agarose concentration.

While \textit{E. coli} showed weak adhesion forces across different LMP agarose concentrations tested, \textit{C. okenii} showed different trend. The bimodal distribution indicates that a few cells bind strongly to the surface macromolecules, while others bind weakly. Similar behaviour had been observed also for \textit{Staphylococcus aureus} on hydrophobic surfaces\cite{maikranz2020different}. Increasing the LMP agarose concentration could provide more anchor points for the \textit{C. okenii} cells, thus increasing the overall adhesion force. However, the exact binding mechanism is currently under investigation, and will be discussed elsewhere.

\begin{figure}[H]
  \centering
  \includegraphics[width=0.9\textwidth]{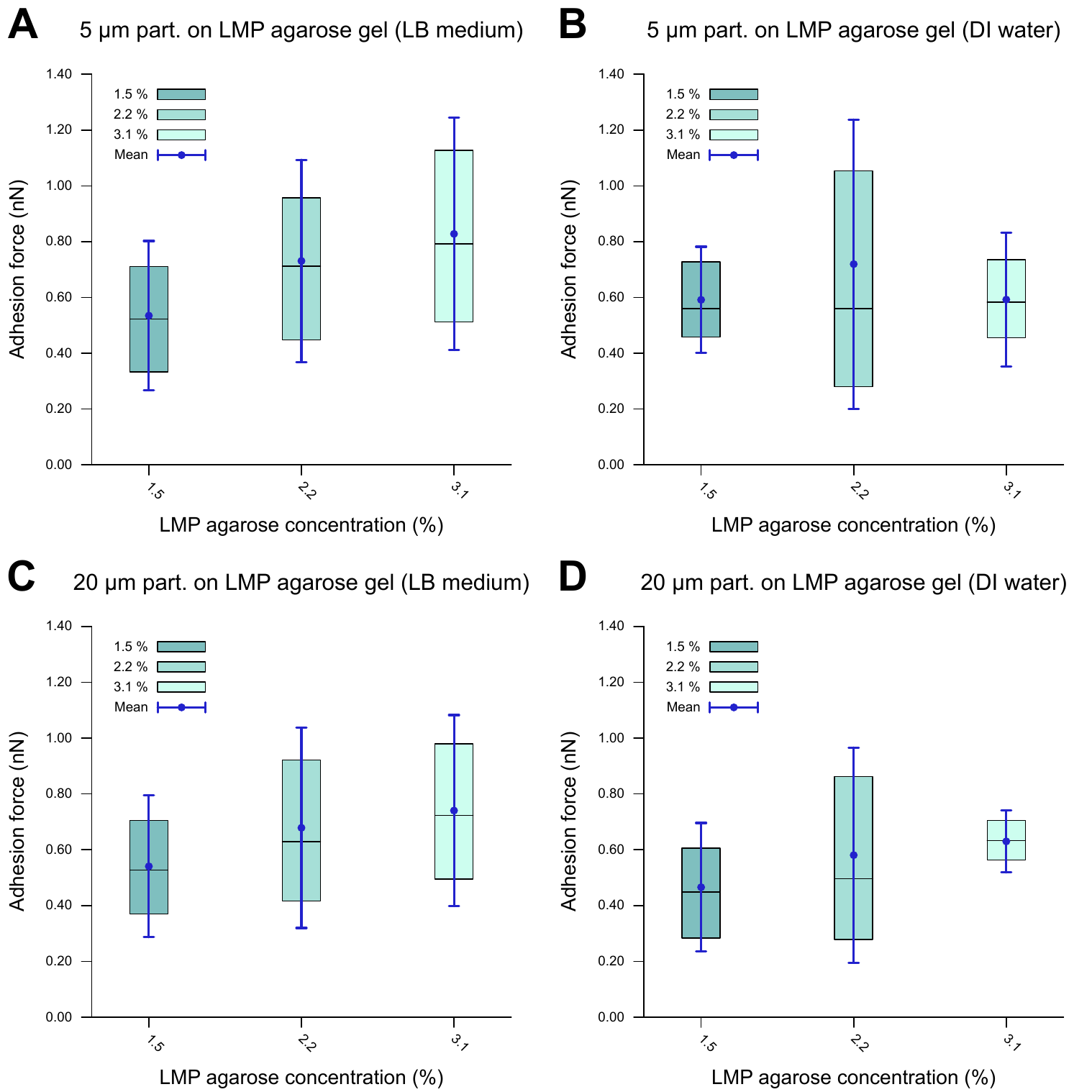}
  \caption{Adhesion forces of polystyrene microparticles towards LMP agarose gel with various concentrations measured in DI water environment. The inner 50~$\%$ of the measured values sorted in ascending order are represented by a coloured area containing the median as a horizontal line. In addition, the average is shown as a point and the standard deviation as an error bar. A) 5~µm particles showing low to medium adhesion towards LB medium based LMP agarose gel with a slight increase in adhesion force as the agarose concentration is increased. B) 5~µm particles show a similar behavior towards DI water based LMP agarose gel as in (A). C,D) 20~µm microparticles show similar behavior towards LB medium based LMP agarose gel (C) as 5~µm microparticles (A) with a similar adhesion force range, and similar behavior towards DI water based LMP agarose gel (D) as 5~µm microparticles (B) with a similar adhesion force range.}
  \label{fig7:microParticlesAdhesion}
\end{figure}
\newpage

\subsection{Adhesion of polystyrene microparticles on LMP agarose gels}
To better understand the adhesion force results and to see if there is an active biological modulation of adhesion, experiments were repeated with passive PS microparticles. As with the bacterial cells, the PS microparticles possessed hydrophobic surfaces. To maintain a comparable size range, microparticles with diameters of 5 µm and 20 µm were used. The system also consisted of an agarose gel based on LB medium as the substrate and DI water as the surrounding medium. To minimize the influence of ions present in the LB medium, these experiments were repeated in a system with agarose gel based on DI water in a DI water environment. The results are shown in Figure~\ref{fig7:microParticlesAdhesion}, while the corresponding values are provided in Table~\ref{tbl4:microparticles}. The adhesion forces fall in the nN range, with marginal increase increase with the agarose concentration. A comparison with the results for \textit{E. coli} and \textit{C. okenii} shows that the adhesion force for PS beads is somewhat higher. Though our results cannot conclusively confirm if \textit{E. coli} actively tune their adhesion with LMP agarose, \textit{C. okenii} exhibits a distinct adhesion trend, suggesting a specific interaction between the substrate and the bacteria that manifests at a specific agarose concentration. This phenomenon aligns with previous findings on the adhesion of \textit{Staphylococcus aureus} \cite{maikranz2020different}, suggesting the presence of diverse active adhesion mechanisms in bacteria.

\begin{table}[]
\centering
  \caption{Adhesion force of polystyrene microparticles on LPM agarose gels.}
  \label{tbl4:microparticles}
  \begin{tabular}{c:cc:cc:c}
    \hline Agarose gel & Particle & Agarose & \multicolumn{2}{c|}{Adhesion force} & Number of\\
    medium & diameter & concentration & Average & Median & measurements\\
    & [µm] & [$\%$] & [kPa] & [kPa] & (\textit{n}) \\

    \hline
    LB medium & 5 & 1.5 & 0.54~($\pm$~0.27) & 0.52 & 902  \\
    LB medium & 5 & 2.2 & 0.73~($\pm$~0.37) & 0.72 & 475 \\
    LB medium & 5 & 3.1 & 0.83~($\pm$~0.41) & 0.80 & 466 \\

    \hline
    LB medium & 20 & 1.5 & 0.45~($\pm$~0.25) & 0.53 & 474  \\
    LB medium & 20 & 2.2 & 0.68~($\pm$~0.36) & 0.63 & 429 \\
    LB medium & 20 & 3.1 & 0.74~($\pm$~0.34) & 0.72 & 256 \\
    \hline
    DI water & 5 & 1.5 & 0.59~($\pm$~0.19) & 0.56 & 389 \\
    DI water & 5 & 2.2 & 0.72~($\pm$~0.52) & 0.56 & 390 \\
    DI water & 5 & 3.1 & 0.59~($\pm$~0.23) & 0.59 & 300 \\

    \hline
    DI water & 20 & 1.5 & 0.47~($\pm$~0.23) & 0.45 & 432 \\
    DI water & 20 & 2.2 & 0.58~($\pm$~0.39) & 0.50 & 319 \\
    DI water & 20 & 3.1 & 0.63~($\pm$~0.12) & 0.64 & 387 \\
    \hline
    \end{tabular}

\end{table}

\newpage

\subsection{Evolution of \textit{C. okenii} adhesion over growth stages}

\begin{figure}[H]
  \centering
\includegraphics[width=0.9\linewidth]{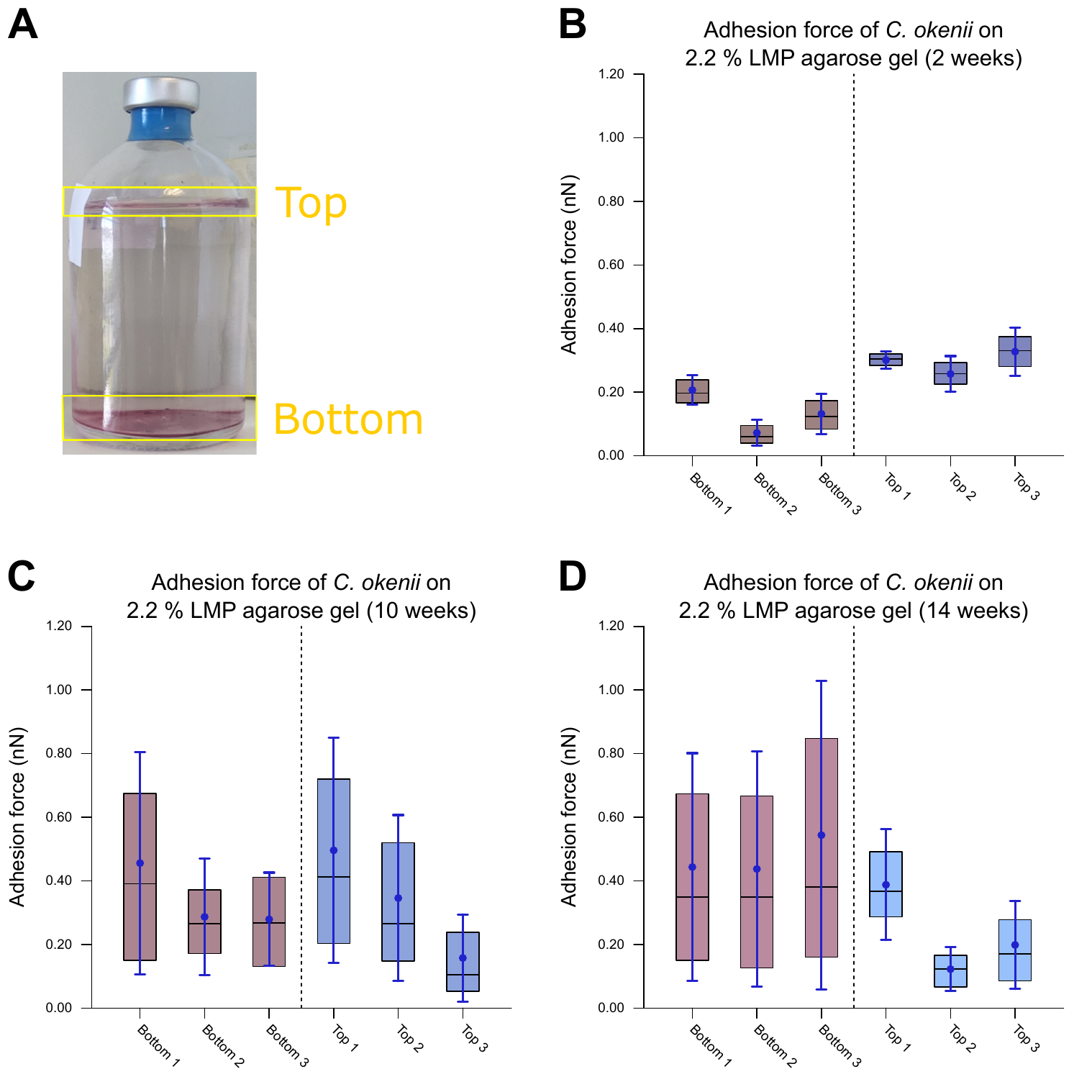}
  \caption{\textit{C. okenii} adhesion force measurements towards 2.2~\% LMP agarose gel. (A) The freshly harvested \textit{C. okenii} cells were stored in a glass jar under anaerobic conditions exposed to room temperature and daylight. A bottom and a top layer of \textit{C. okenii} formed within the vessel. For each chosen time point, three samples of the top layer and three samples of the bottom layer were measured. For B, C and D, the inner 50~$\%$ of the measured values sorted in ascending order are represented by a coloured area containing the median as a horizontal line. In addition, the average is shown as a point and the standard deviation as an error bar. (B) Adhesion force results of bottom and top layer cells 2 weeks after removal from natural habitat, showing small adhesion only. (C) Adhesion force results of bottom and top layer cells 10 weeks after removal from natural habitat, Showing increased adhesion force range for all samples. (D) Adhesion force results of bottom and top layer cells 14 weeks after removal from natural habitat, showing large adhesion force range especially for bottom cells.}
  \label{fig8:cOkeniiFreshDetails}

\end{figure}

As a final step of our experiments, we measured the evolution of the \textit{C. okenii} adhesion forces during the course of laboratory inoculation. Recent report by the authors have demonstrated that, under laboratory conditions, \textit{C. okenii} undergoes a lifeform shift from a free-living motile to biofilm forming sessile phenotypes \cite{di2023synergistic}. Since bacterial adhesion to surfaces is central to biofilm formation, the final measurements are motivated by the question: Does the lifeform shift of \textit{C. okenii} correspond to changes in their cell-substrate adhesion properties? Regardless of the exact mechanism of adhesion between \textit{C. okenii} and the LMP agarose substrate, it is intriguing that \textit{C. okenii} populations elicit wek and strong adherent phenotypes at higher substrate stiffness. Relative to the planktonic states with suppressed cell-surface adhesion properties, transition to sessile lifeforms could be triggered by amplified adhesion mechanisms (for instance, via secretion of exopolymeric substances under environmental stressors), driving cellular flocculation and surface attachment, ultimately leading to sessile lifeforms\cite{sengupta2023planktonic}. Thus, starting with a fresh sample of planktonic \textit{C. okenii} isolated from their natural lake habitat, we analysed their cell-substrate adhesion forces, and compared them with lab-grown cultures over a period of 14 weeks (late stationary phase, \cite{di2023synergistic}). Macroscopically, we observed two distinct \textit{C. okenii} layers in the inoculation bottle: a bottom layer attached to the floor of the glass bottom, and a top suspended layer at the air-water interface, as visualized and labeled in Figure~\ref{fig8:cOkeniiFreshDetails}.

\begin{figure}[H]
  \centering
  \includegraphics[width=\linewidth]{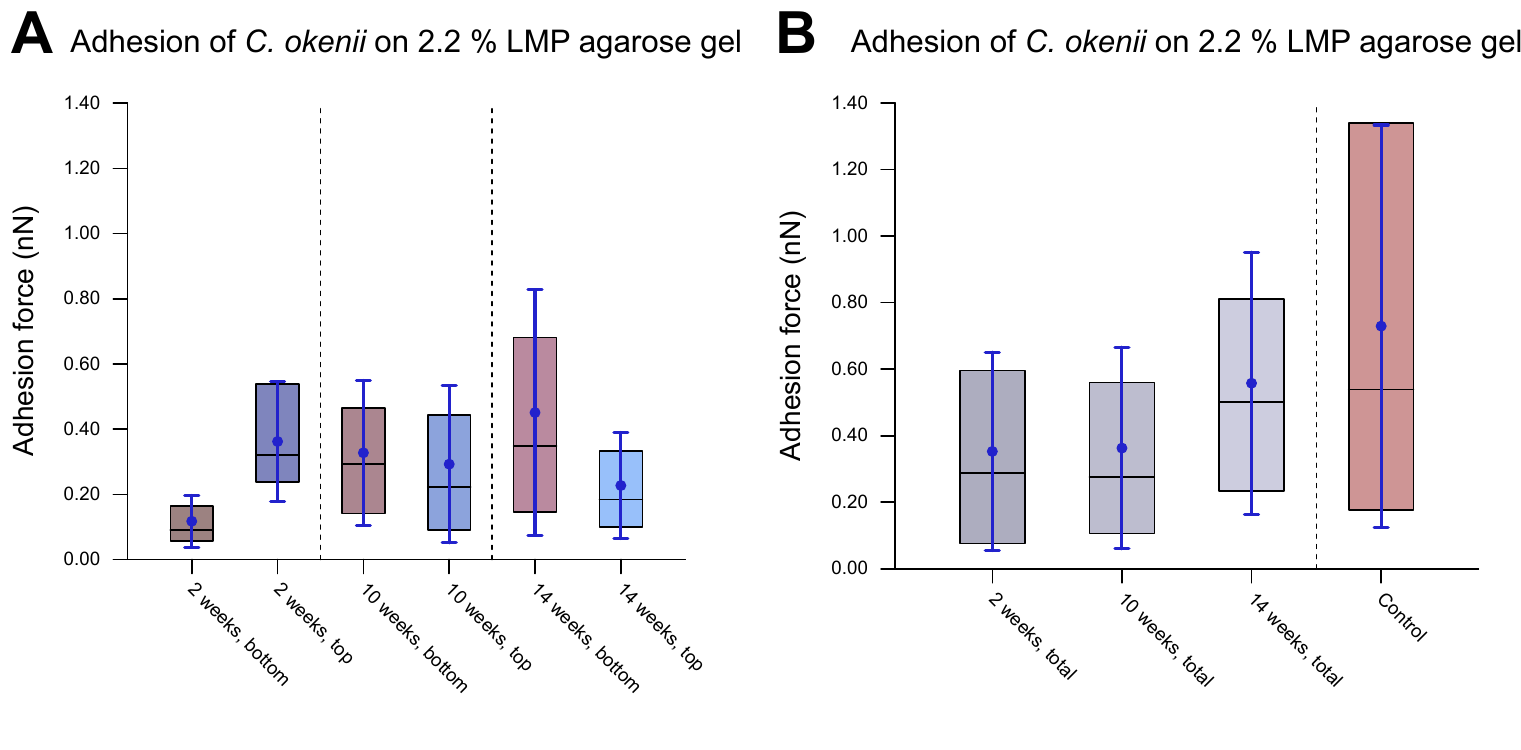}
  \caption{
  Evolution of \textit{C. okenii} cell adhesion towards 2.2~\% LMP agarose gel over time in laboratory conditions. The inner 50~$\%$ of the measured values sorted in ascending order are represented by a coloured area containing the median as a horizontal line. In addition, the average is shown as a point and the standard deviation as an error bar. The plots show the evaluated adhesion force data 2, 10 and 14 weeks after transfer from lake to laboratory. A) shows separated data of bottom and top cells. As can be seen, the top cells initially show slightly higher adhesion than the bottom cells. B) shows the combined bottom/top cells data in function of time. For comparison, control cells of 8th generation in laboratory are shown on the right. The increase of adhesion with domestication time indicates the growing ability of cells to form biofilms. Furthermore, the growing adhesion range is an indication for different development stage either towards biofilm formation or within biofilm circle.
}
  \label{fig9:cOkeniiCellAdhesion}
\end{figure}

Figure~\ref{fig8:cOkeniiFreshDetails}B-D illustrates the adhesion values of both top and bottom cells to a 2.2~\% agarose gel substrate over time since their removal from their natural habitat and maintenance under laboratory conditions. The combination of top and bottom cells from each time period is shown in Figure~\ref{fig9:cOkeniiCellAdhesion}, including data for populations grown into late stationary phase (Figure~\ref{fig9:cOkeniiCellAdhesion}B). Table~\ref{tbl5:cOkeniiFreshDetails} lists corresponding values.

After two weeks of laboratory inoculation, i.e., when the \textit{C. okenii} population is in its mid-exponential phase, the cell adhesion forces fell within a narrow range, with \textit{top} cells exhibiting higher adhesion ($0.36 \pm 0.18$ nN), relative to their \textit{bottom} counterparts ($0.12 \pm 0.07$ nN). The scenario changes after ten weeks, when the \textit{C. okenii} population enters stationary phase: the bottom sub-populations have an enhanced adhesion, with an overall increase in the diversity of the adherent phenotypes (as indicated by the high standard deviations in our measurements, see Table~\ref{tbl5:cOkeniiFreshDetails}). The difference between median values of upper and lower cells decreased. Finally, once the population reached late stationary phase (at 14 weeks), the bottom population is now considerably more adherent, with a larger diversity of the adherent phenotypes. The top population however maintains its adhesion. A closer examination of the separate measurements for bottom and top cells revealed a progressive evolution of adherent characteristics of the bottom cells, whereas the adhesion properties of the top cells appear to stabilize over the same period of time. Thus, not only do the mean and median adhesion forces increase, the entire range of adhesion forces, including the maximum adhesion force, enhance over time, accompanied by an overall decrease in the number of \textit{top} cells. 

\begin{table}[H]
\centering
  \caption{Adhesion force of \textit{C. okenii} to LMP agarose gel across growth stages}
  \label{tbl5:cOkeniiFreshDetails}
  \begin{tabular}{c:c:cc:c}
    \hline
    Time since & Extraction & \multicolumn{2}{c:}{Adhesion force} & Number of \\
    inoculation & point & Average & Median & measurements \\
    \multicolumn{1}{c:}{{[weeks]}} & & [nN] & [nN] & (\textit{n}) \\
    \hline
    2 & Bottom & 0.12~($\pm$~0.07) & 0.09 & 1479 \\
    2 & Top & 0.36~($\pm$~0.18) & 0.32 & 1310 \\
    10 & Bottom & 0.33~($\pm$~0.23) & 0.29 & 1940 \\
    10 & Top & 0.23~($\pm$~0.24) & 0.22 & 2042 \\
    14 & Bottom & 0.45~($\pm$~0.37) & 0.35 & 2338 \\
    14 & Top & 0.23~($\pm$~0.16) & 0.19 & 1922 \\
    \hline
  \end{tabular}
\end{table}

\newpage

\section{Discussions and Conclusion}

In this work, we present a Force Distance Spectroscopic approach to simultaneously measure the stiffness and adhesion of low-melting-point agarose gel, both in its native state as well as in the context of cell-surface adhesion of bacterial species. Our results show that with increasing agarose concentration, the substrate stiffness goes up, however, remains below the typical value of normal melting agarose gels. The Young's moduli of LMP agarose gels increase from approximately 20~kPa to around 100~kPa as the LMP agarose concentration is doubled from 1.5~\% to 3~\%. The reported range of Young's moduli falls between ultralow-melting-point agarose and the normal-melting-point agarose, and thus offers a well-suited alternative to existing hydrogels used in biomedical research and engineering. No significant difference was observed depending on the choice of the liquid environments (LB versus DI water) within which the measurements were conducted. The stiffness of LMP agarose gel lies in the range of soft tissues such as muscle, epithelial and neural tissues\cite{ji2021recent}, making it potential material for organ-on-a-chip tissue engineering applications.  

FDS experiments with \textit{E. coli} and \textit{C. okenii} bacteria on LMP agarose gel with three different concentrations revealed a stiffness-dependent variation of the cell-surface adhesion. While \textit{E. coli} showed marginal increment of the adhesion with increasing agarose content, the attachment of \textit{C. okenii} showed two different adherent phenotypes: one with adhesion force in the same range as observed for \textit{E. coli} (weak adhesion); and a second sub-population of strongly adherent cells as well as increased adhesion with the LMP agarose concentration. Previous works with diverse species and substrates have indicated inconsistent adhesion forces, underscoring the possibility of cofounding factors, beyond stiffness, to contribute to the dynamics of bacteria-substrate adhesion. For example, Thio \textit{et al.} determined the adhesion forces of \textit{E. coli} bacteria towards polystyrene and different polyamides to be in the range of 2.9~nN to 9.7~nN\cite{thio2008quantification}, while Wang \textit{et al.} reported that \textit{Staphylococcus aureus} adhesion on polyacrylamide hydrogels decrease by three orders with increasing substrate stiffness\cite{wang_interactions_2016}. While \textit{E. coli} and \textit{P. aeruginosa} show a similar adhesion trend on polydimethylsiloxane substrate (0.1~MPa to 2.6~MPa)\cite{song2014stiffness}, \textit{E. coli} and \textit{S. aureus} show an increased adhesion to stiffer poly(ethylene glycol) dimethacrylate hydrogels, relative to softer ones\cite{kolewe_fewer_2015,kolewe_bacterial_2018}. This is also supported by Francius \textit{et al.}\cite{francius2021impacts} who observed a linearly increasing adhesion dependence of different \textit{E. coli} strains on poly(allylamine hydrochloride)/hyaluronic acid hydrogels stiffness between 20~kPa and 700~kPa. Guégan \textit{et al.}\cite{guegan2014alteration} measured the effect of the stiffness of agarose gel of two different concentrations (0.75~\% and 3.0~\%) on the adhesion capability of gram-negative \textit{Pseudoalteromonas}~sp.~D41 and gram-positive \textit{Bacillus}~sp.~4J6 using retention assays. While \textit{Pseudoalteromonas}~sp.~D41 showed higher adhesion for softer agarose substrate, adhesion of \textit{Bacillus}~sp.~4J6 increased with increasing stiffness.

\begin{figure}[H]
  \centering
\includegraphics[width=0.9\linewidth]{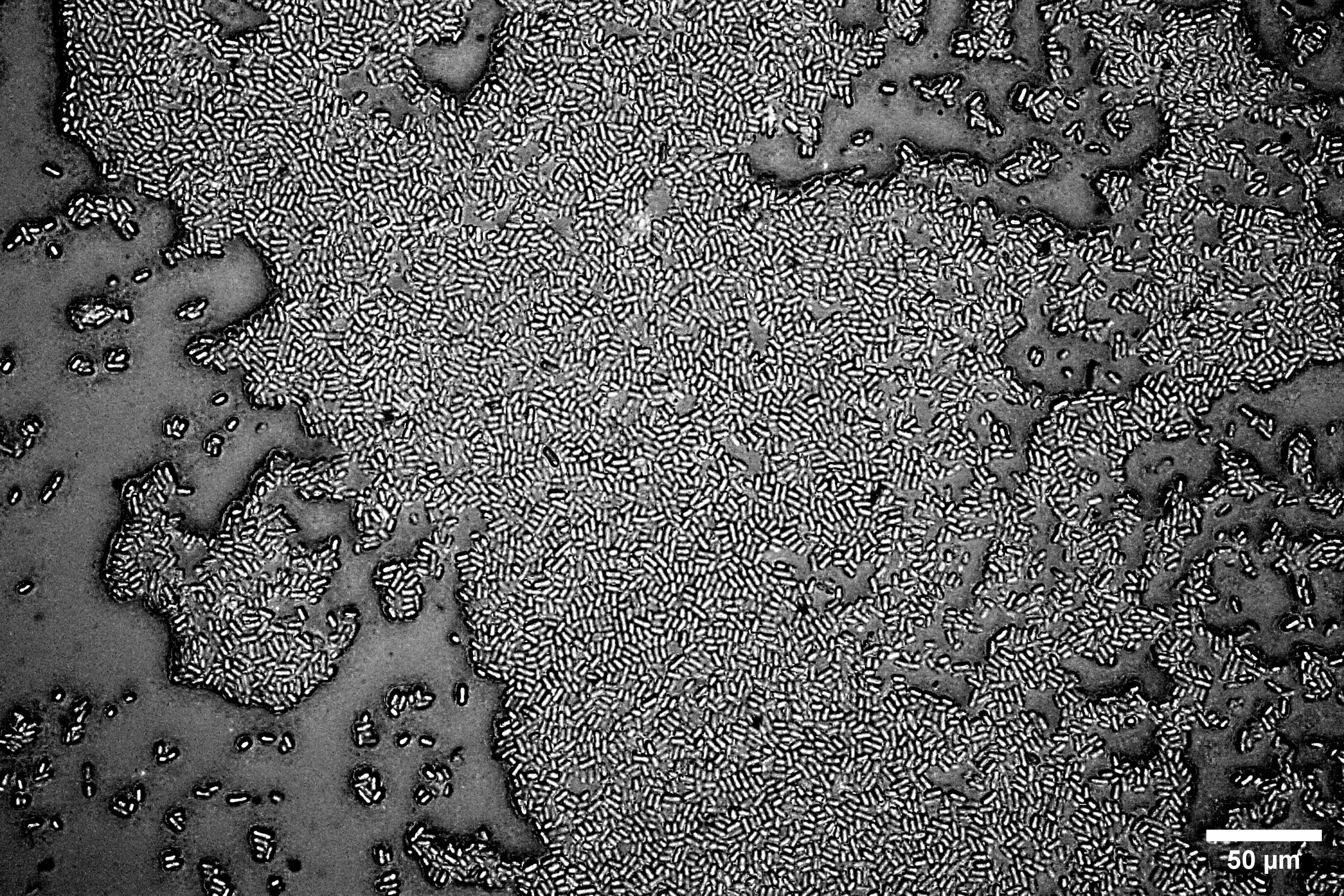}
  \caption{\textit{C. okenii} biofilm on 2.2~\% LMP agarose gel for a 14 week old culture. The cells form a monolayer on the underlying LMP agarose substrate.}
  \label{fig10:cOkeniiBiofilm}
  
\end{figure}

The mechanism of this adhesion process depends on the properties of the bacteria, substrate and the local environment\cite{berne2018bacterial,gordon_bacterial_2019,kimkes2020bacteria}. For \textit{C. okenii}, the larger contact area of surface polymeric molecules on the substrate, as well as its surface chemistry, may promote the observed adhesion and the adherent diversity on the agarose gel substrates. Our experiments over the physiological growth stages of \textit{C. okenii} indicate a progressive increase in adhesion (on 2.2~\% agarose gel) over time, supporting the lifeform shift from planktonic to biofilm state\cite{di2023synergistic}. The \textit{C. okenii} biofilm formation could be observed stably after 14 weeks of inoculation, as shown in Figure~\ref{fig10:cOkeniiBiofilm}. As reported previously by Di Nezio \textit{et al.}\cite{di2023synergistic}, such shifts in lifeforms trigerred by domestication of species, occur synergistically across multiple phenotypic traits, including morphology, cell density, intracellular attributes, and alteration of motility. The cell-substrate adhesion data supports the previously reported shift in the \textit{C. okenii} lifeform, occurring over a period of weeks after laboratory inoculation. The phenotypic diversity of the adherent cells increase over time, furthermore suggesting an adaptive transformation which could allow the domesticated cells to optimise surface attachment once they are inoculated over extended periods under laboratory conditions. Alteration of cell adhesion following domestication has been reported also for \textit{Saccharomyces cerevisiae}\cite{kuthan2003domestication}, wherein such changes were mediated by modified expression of genes related to cell-cell adhesion and the production of EPS \cite{kuthan2003domestication,Barua2016Molecular}. Similar pathways could be targeted in follow up studies to delineate how planktonic \textit{C. okenii} undergo lifeform shifts, specifically in the context of stiffness- and physiology-dependent cell adhesion.  

\newpage

\section*{Acknowledgments}
 We thank F. Di Nezio and N. Storelli for sharing \textit{Chromatium okenii} cells and valuable discussions on their growth and physiology. We gratefully acknowledge the support from the Institute for Advanced Studies, University of Luxembourg (AUDACITY Grant: IAS-20/CAMEOS to A.S.) and a Human Frontier Science Program Cross Disciplinary Fellowship (LT 00230/2021-C to G.R.). A.S. thanks Luxembourg National Research Fund for the ATTRACT Investigator Grant (A17/MS/ 11572821/MBRACE) and a CORE Grant (C19/MS/13719464/TOPOFLUME/Sengupta) for supporting this work.

\section*{Conflict of Interest}
The authors declare no conflict of interest.

\section*{Author contributions}
Conceptualization, planning, administration, and supervision: A.S. Methodology: R.R., G.R. and A.S. Investigation, data and statistical analysis: R.R. with inputs from A.S. Writing: R.R. and A.S.

\section*{Data Availability Statement}
All data and supporting materials are included in the manuscript. Any additional supporting material for this study can be obtained from the corresponding author upon reasonable request.

\newpage

\bibliography{scibib}

\bibliographystyle{Science}


\clearpage

\end{document}